\documentclass[prl,twocolumn,showpacs,superscriptaddress,10pt]{revtex4-1}
\usepackage {graphicx} 
\usepackage {amssymb} 
\usepackage {amsmath}
\usepackage {mathrsfs}

\newcommand{\vect}[1]{\mathrm{\mathbf{#1}}} 
\DeclareMathOperator\erf{erf}
\DeclareMathOperator\sinc{sinc}
\DeclareMathOperator\dsinc{dsinc}
\DeclareMathOperator\ddsinc{ddsinc}

\newcommand{\reffig}[1]{Fig.~\ref{#1}}
\newcommand{\refeq}[1]{Eq.~(\ref{#1})}
\newcommand{\refeqs}[2]{Eqs.~(\ref{#1})-(\ref{#2})}

\begin{document}

\title{Boosting terahertz generation in laser-field ionized gases using a
  sawtooth wave shape}

\author{P. Gonz{\'a}lez de Alaiza Mart{\'i}nez}
\affiliation{CEA-DAM, DIF, F-91297 Arpajon, France}

\author{I. Babushkin}
 \affiliation{ Institute of  Quantum Optics, Leibniz University Hannover,
   Welfengarten 1 30167, Hannover, Germany}

\author{L. Berg\'e}
\affiliation{CEA-DAM, DIF, F-91297 Arpajon, France}

\author{S. Skupin}
\affiliation{Univ. de Bordeaux - CNRS - CEA, Centre Lasers Intenses et Applications, UMR 5107, 33405 Talence, France}

\author{E. Cabrera-Granado}
\affiliation{Facultad de \'Optica y Optometr\'ia, Universidad Complutense de Madrid, 28037,
Madrid, Spain}

\author{C. K\"ohler}
\affiliation{Institute for Scientific Computing, TU Dresden, 01062 Dresden, Germany}

\author{U. Morgner}
 \affiliation{ Institute of  Quantum Optics, Leibniz University Hannover,
   Welfengarten 1 30167, Hannover, Germany}

\author{A. Husakou}
\affiliation{Max-Born-Institut f\"ur Nichtlineare Optik und Kurzzeitspektroskopie, 12489
Berlin, Germany}

\author{J. Herrmann}
\affiliation{Max-Born-Institut f\"ur Nichtlineare Optik und Kurzzeitspektroskopie, 12489
Berlin, Germany}

\date{\today}


\begin{abstract}
  Broadband ultrashort terahertz (THz) 
  pulses can be produced using plasma generation in a
  noble gas ionized by femtosecond two-color
  pulses.  Here we 
  demonstrate  that, by using multiple-frequency
  laser  pulses, one can
  obtain a waveform which optimizes the free electron
  trajectories in such a way that they reach the
  highest velocity at the electric field extrema. This
  allows to increase the THz conversion efficiency to the percent level, 
    an unprecedented performance for THz generation in gases.
   Besides the analytical study of THz generation
     using a local current model, we
    perform comprehensive 3D simulations accounting for propagation
  effects which confirm this prediction. Our
  results show that THz conversion via
   tunnel ionization
  can be greatly improved with well-designed multicolor
  pulses.  
\end{abstract}

\maketitle

Ultrashort pulses in the terahertz (THz) range (from $\sim 0.1$ to $\sim 30$
THz) are extremely important for various
time-resolved studies in molecular physics, chemistry, material
sciences, and security
applications~\cite{mittleman02,kress:nphy:2:327,pickwell:jpd:39:R301,Tonouchi:np:1:97,
  marx:lfw:43:44,chan:rpp:70:1325,hoffmann:jpd:44:083001,woerner:njp:15:025039}.  The
THz frequency interval lies in between the accessible range of
electronic (microwave) and optical (infrared) technologies, often
referred to as the THz gap.  To access this spectral range, recent
investigations proposed to use nonlinear  processes in gases ionized by fs laser pulses, which provides higher breakdown thresholds, broader spectral ranges and better tunability in the THz and mid-infrared (MIR)  ranges. Field strengths of THz pulses generated by two-color optical pulses are  typically  limited to the range below 1 MV/cm or pulse energies below 100 nJ \cite{Cook:ol:25:1210,Kress:ol:29:1120,Bartel:ol:30:2805,kim:oe:15:4577,Kim:np:2:605,Roskos:lpr:1:349,Reimann:rpp:70:1597}.
The highest THz energy of 5 $\mu$J was obtained in \cite{Kim:np:2:605} with conversion efficiency of
  10$^{-4}$.  The physics
behind this so-called photocurrent mechanism can be understood by
 the peculiarities of tunnel ionization and
subsequent dynamics of free electrons in the
field~\cite{kim:oe:15:4577,Kim:np:2:605,Babushkin:oe:18:9658,Babushkin:njp:13:123029,Koehler:ol:36:3166,berge:prl:110:073901}. It
was 
shown~\cite{Babushkin:oe:18:9658,Babushkin:njp:13:123029,berge:prl:110:073901}
that laser-driven THz emission is related to interactions over two
different time scales:
the attosecond  sub-cycle
  dynamics of tunnel ionization and the femtosecond 
 pump pulse dynamics. Free electrons,
extracted from atoms by ionization in sharp attosecond-long
steps~\cite{Uiberacker:nat:446:627,verhoef:prl:104:163904,Babushkin:njp:13:123029},
are accelerated in the laser field and create a net macroscopic current that contains low frequency
components. This current is responsible for the observed THz emission.

In Refs.~\cite{Babushkin:oe:18:9658,Babushkin:njp:13:123029,berge:prl:110:073901} the spectral shape and
energy of the THz pulses generated by the above mechanism have been explained by
 analogy with linear
diffraction theory, where ionization events play the role of slits
in a ``temporal diffraction grating``. The THz radiation then appears as a zero-diffraction
order peak in the corresponding diffractive pattern. 
Its amplitude is impacted by each ionization event and is determined not only by the pump field near 
the ionization instants, but also by  the
whole pump waveform. This non-instantaneous dependence suggests that,
by optimizing the pump field shape, one may achieve higher THz yields
for a given pulse energy. 

In this article, we exploit degrees of freedom given by appropriately
 chosen pump waveforms to increase the THz conversion
 efficiency. We show that the fields with a sawtooth-like temporal shape
 do promote the highest THz signals triggered by 
tunnel-induced photocurrents. We  predict, by means of a local theory,
that sawtooth waveforms can in principle increase the THz efficiency
by up to two orders of magnitude, 
compared to a standard two-color pulse. 
Boosting the THz yield is confirmed
through comprehensive 3D computations that take all propagation
effects into account. 
Selecting the first four
Fourier harmonics of the sawtooth waveform already guarantees an
impressive  
increase of the THz  energy up to
  5 $\mu$J, similar to the record value in  \cite{Kim:np:2:605} but with 100 times smaller total pump pulse energy.

We start with the local current (LC)
approximation~\cite{Babushkin:njp:13:123029}, which neglects propagation effects. The free electron density $\rho(t)$ and current $J(t)$ are governed by
\begin{align}
\label{eq:rho}
\frac{\partial \rho(t)}{\partial t} & = W[E(t)]\left[\rho_0 -\rho(t) \right], \\
\label{eq:j}
\frac{\partial J(t)}{\partial t} & = \frac{q^2}{m}\rho(t)E(t)-\frac{J(t)}{\tau_c},
\end{align}
where $E(t)$ is the pump field;
\begin{equation}\label{eq:tunnel}
W[E(t)] = \frac{\alpha}{|E|}\exp\left[-\frac{\beta}{|E|}\right]
\end{equation}
is the instantaneous tunnel ionization
rate~\cite{Roskos:lpr:1:349}, $\rho_{0}$ is the density of neutral atoms, $q$ and $m$ are the charge and mass of electron,
and $\tau_c$ is the current decay time due to collisions. 
In \refeq{eq:tunnel}, $\alpha = 4\omega_{a}(r_{H})^{5/2} E_{a}$ and $ \beta = 2(r_{H})^{3/2} E_{a}/3$ depend on the ratio of the ionization potential $U_i$ of the considered gas over the hydrogen ionization potential, $r_H=U_i/U_{H}$, while
$E_a=m^2q^5/(4\pi\varepsilon_0)^3\hbar^4$ and $\omega
  _{a}=mq^{4}/(4 \pi \epsilon_0)^2 \hbar ^{3}$.
The THz field $E_{\rm THz}(t)$ is generated by the free electrons %
created  by tunnel ionization and then accelerated
in the pump field. That is, assuming a small size of the plasma spot, $E_{\rm THz}(t) \approx g
\partial_t J(t)$, where $g$ is a geometrical factor depending on
the position of the observer~\cite{Babushkin:njp:13:123029}.

Ionization mostly happens near the extrema of $E(t)$. In the following, we number their corresponding instants consecutively as $t_1,t_2,t_3,\ldots t_n$. Thus, 
the electron density and current [\refeqs{eq:rho}{eq:j}] can be
approximated as follows (see~\cite{Babushkin:njp:13:123029} for details):
\begin{align}
\label{eq:rhomodel}
\rho(t) & \simeq \sum_n\delta\rho_n H_n(t-t_n), \\
\label{eq:jmodel}
J(t) & \simeq J_A(t) + J_B(t),\\
\label{eq:JA}
J_A(t) & = \sum_n q \delta\rho_n  v_f(t) H_n(t-t_n), \\
\label{eq:JB}
J_B(t) & = - \sum_n q \delta\rho_n e^{-\frac{t-t_n}{\tau_c}}v_f(t_n) H_n(t-t_n),
\end{align}
where  $v_f(t)$
is the free electron velocity  given by
\begin{equation}
\label{eq:vf}
v_f(t) = \frac{q}{m}e^{-\frac{t}{\tau_c}}\int_{-\infty}^tE(t')e^{\frac{t'}{\tau_c}}\,dt'.
\end{equation}
The  quasi-step function used in \refeqs{eq:rhomodel}{eq:JB}
is $H_n(t)=\frac{1}{2}[1+ \erf(t/\tau_n)]$, where $\tau_n$ is the width of
the $n$th ionization event and the density jump at $t=t_n$, $\delta\rho_n$, is expressed as (see supplemental material)
\begin{equation}
\label{eq:rhon}
\delta\rho_n \simeq \rho_0 \epsilon_n \left(1-e^{-\sqrt{\pi}W[E(t_n)]\tau_n} \right),
\end{equation}
with $\epsilon_1 = 1$ and $\epsilon_n = e^{-\sqrt{\pi}
  \sum_{j=1}^{n-1}W[E(t_j)]\tau_j}$ for $n>1$.  

\begin{figure}
\includegraphics[width=\columnwidth]{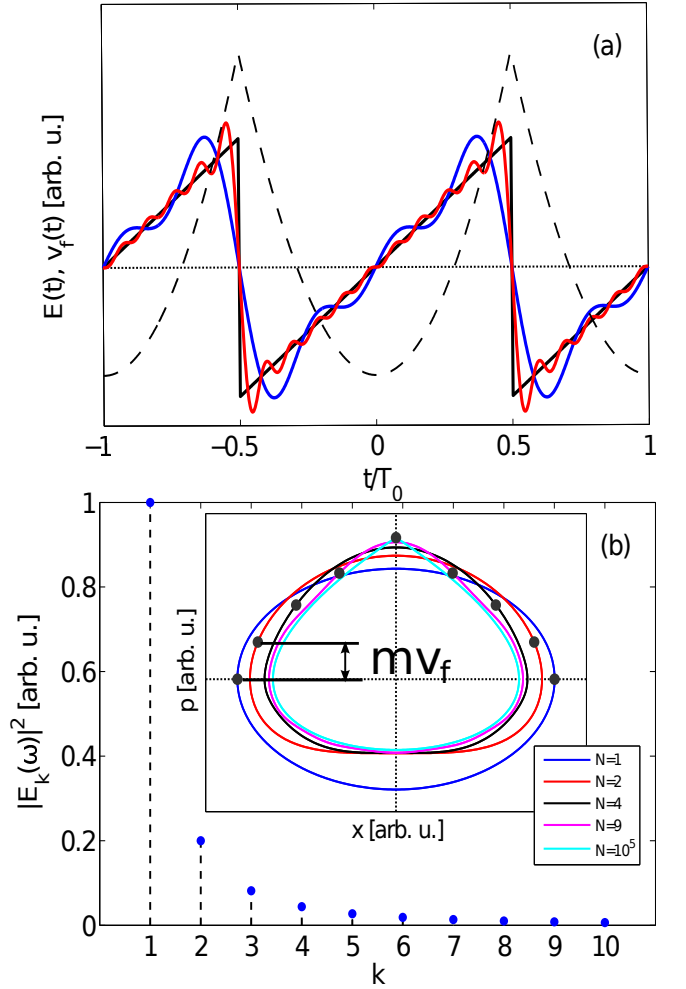}
\caption{(a) Sawtooth waveform $E(t)$ (black thick solid line)
  having the maxima of $v_f(t)$ (black  dashed line) at the same
  instants as the extrema of $E(t)$. Colored solid curves show the three-color (blue curve) and
  ten-color (red curve) approximations to the ideal sawtooth
  shape. (b) Spectrum of the sawtooth waveform containing all
  harmonics of $\omega_0$ with intensities falling down as
  $1/k^2$. Inset shows the trajectories of free electrons in the phase
  space $(p,x)$ for an increasing number of colors. Dots locate the
  maxima of $\left| E(t) \right|$;
    for the 2-color case, the value of $m v_f$ at those maxima is
    exemplified.}
  \label{fig1}
\end{figure}

Moreover, it is
possible to estimate analytically the spectrum of the THz radiation generated by  the current
components $J_A$ and $J_B$. We obtain in Fourier space (see supplementary material for a
detailed derivation)
\begin{subequations}
\label{eq:freqJAB}
\begin{align}
\mathcal{F}[\partial_t J_B](\omega) & \approx
  \frac{-q}{\sqrt{2\pi}}\sum_n\delta\rho_n
  v_f(t_n)e^{it_n\omega}\frac{\omega}{i/\tau_c+\omega},\, \label{eq:freqJB}\\
  \mathcal{F}[\partial_t J_A](\omega) & \approx \frac{-iq^2}{\sqrt{2\pi} m} \sum_n 
 \sum_{k=1}^N \delta\rho_n E_k(t_n) \frac{\omega}{k^2\omega_0^2}. \label{eq:freqJA}
\end{align}
\end{subequations}
Here to evaluate $J_A(t)$, we considered multi-color pulses of the form
\begin{equation}
\label{eq:ea}
E(t) = \sum_{k=1}^N E_k(t)=  \sum_{k=1}^N \mathcal{E}_k(t) a_k \cos(k \omega_0 t + \phi_k),
\end{equation}
where $\mathcal{E}_k(t)$, $a_k$ and $\phi_k$ are the envelope with duration $\tau_k$, 
relative amplitude and phase of the $k$th harmonic, respectively;  $\omega_0$ is the fundamental frequency and we assumed long pulses $\omega_0 \tau_k \gg 1$.
 For
a near-infrared pump, the contribution $J_B$ dominates over $J_A$
\cite{2014arXiv1407.6893C} in the THz spectral range, mainly because the
$E_k(t_n)$ are not sign definite and the summands in 
\refeq{eq:freqJA} mostly cancel each other. 
We therefore neglect $J_A$ in the coming analysis. We
  also assume  henceforth  Gaussian envelopes $\mathcal{E}_k(t) = E_0 e^{-t^2/\tau^2}$ with
amplitude $E_0$ and duration $\tau$ being identical for all
colors. 

The energy in the THz part of the spectrum below a certain cut-off frequency $\omega_{\rm{co}} \gg 1/\tau_c$ can be thus
   estimated by the local THz yield as 
$U_{\rm THz} \propto
  \int\limits_0^{\omega_{\rm{co}}} |\omega \hat{J}_B(\omega)|^2 d\omega
  \propto \left[\sum\limits_n \delta\rho_nv_f(t_n)\right]^2.$ 
 Obviously,  to maximize $U_{\rm THz}$ one may not only try to
  increase the ionization yield $\delta \rho_n$, but also the free electron
  velocity $v_f(t_n)$. Thus, maximizing the THz energy  requires that
$v_f(t)$ reaches its extreme values at the discrete instants $t_n$,
and that all $v_f(t_n)$ have the same sign. According to
\refeq{eq:vf}, $v_f(t)$ attains local extrema when $E(t)$ changes its
sign. So the only way to achieve maxima of both $\left| E(t) \right|$
and $v_f(t)$ at the same instants is to have a discontinuous electric
field. This condition is fulfilled by the sawtooth waveform, which can
be obtained by setting $a_k=1/k$ and $\phi_k=(-1)^k\pi/2$ in \refeq{eq:ea}.
The sawtooth waveform and $v_f(t)$ are  illustrated by Fig.~\ref{fig1}(a)
in the limit of negligible free electron collisions, $\tau_c \rightarrow +\infty$. Figure~\ref{fig1}(a) also
shows that the velocities $v_f(t_n)$ are sign
definite.

Practically, as an infinite number of harmonics in the pump is not achievable, we should employ a finite number $N$ of
colors. As shown in \reffig{fig1}(a), the sawtooth shape is fairly
well reproduced for $N \geq 3$, which is confirmed by the sawtooth
spectrum of Fig. \ref{fig1}(b). Inset of this figure details the free electron phase space 
$p(t) \propto \int_{-\infty}^t E(t') dt' \sim mv_f(t)$ versus  $x(t)$, 
as well as the maximum values of $\left| E(t) \right|$ (see
dots). One can see that the sign-definite value of $p(t)$ at the field
extrema indeed increases with the number of colors, thus
 increasing $U_{\rm THz}$. 

In the following, we fix
 $\omega_0$ corresponding to the
wavelength $\lambda_0=1600$~nm, a choice  clearly advantageous
over the usual  one $\lambda_0 =800$~nm, because  more harmonics are accessible in practice.  In
  particular, the first four harmonics are $\lambda_0/2=800$~nm,
$\lambda_0/3=533$~nm, and $\lambda_0/4=400$~nm. All these frequencies
can be produced from a  $800$~nm  femtosecond laser source using,
for instance, optical parametric amplification to obtain $\lambda_0$ and $\lambda_0/3$ and
 frequency doubling to obtain
  $\lambda_0/4$. In contrast, for $\lambda_0=800$~nm  the fourth harmonic at
200~nm lies already in the UV and is not so easy to
produce. Throughout the paper, we consider  argon at 1~atm pressure and assume a Gaussian pulse envelope with 40~fs FWHM duration ($\tau=34$~fs).

Let us first check our analytical predictions by computing numerically
the local THz yield $U_{\rm{THz}}$ in the frequency range $\nu \equiv \omega/2\pi
<\nu_{\rm{co}}\equiv \omega_{\rm co}/2\pi=100$~THz from the LC
model. For a given multi-color pulse [\refeq{eq:ea}] we compute the
current $J(t)$ according to~\refeqs{eq:rho}{eq:tunnel} and evaluate the
local THz yield in Fourier space $U_{\rm{THz}} \propto
\int\limits_0^{\nu_{\rm{co}}} |\nu \hat{J}(\nu)|^2
d\nu$~\footnote{Note that here we take into account the total current
  $J$, without neglecting $J_A$.}. 
 From our coming 3D simulations
accounting for propagation effects we know that for given pulse
energy and focusing conditions the ionization yield for different
multi-color configurations is almost the same [see
Fig.~\ref{fig3}(d)]. This is understandable, because free electrons
have a strong defocusing effect and balance the intensity growth in
the focal region, similar to the well-known intensity clamping in
femtosecond
filaments~\cite{Berge:rpp:70:1633}.
A reasonable strategy is thus to
compare the local THz yield from pulses producing a constant
ionization level controlled by $\rho(t \rightarrow +\infty) = \rho^{\rm{max}}$ in \refeq{eq:rho}.  

\begin{figure}
  \includegraphics[width=\columnwidth]{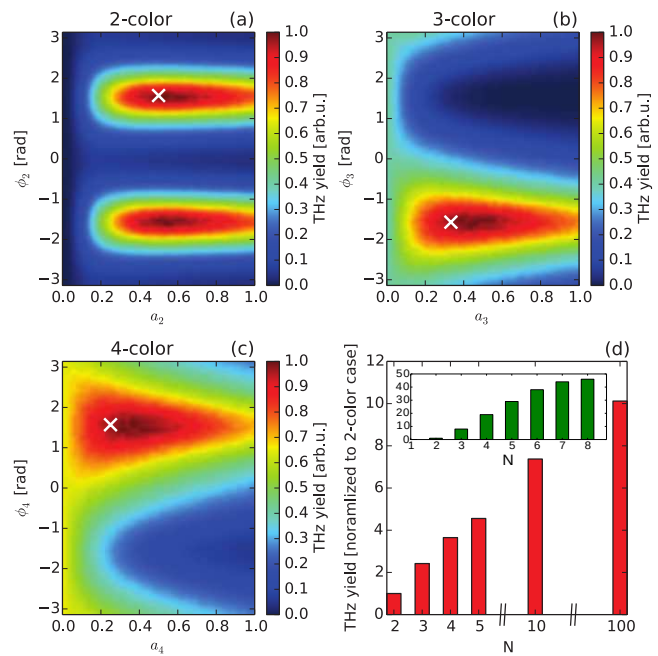}
  \caption{(a) Dependency of local THz yield on $a_2,\phi_2$ for a
    $N=2$ color field \refeq{eq:ea}. (b) Same for $a_3,\phi_3$ and
    $N=3$. (c) Same for $a_4,\phi_4$ and $N=4$. Relative amplitude and
    phases of the lower harmonics in (a)-(c) are fixed according to
    the (optimum) sawtooth shape, i.e., $a_k=1/k$, $\phi_k=(-1)^k\pi/2$. The
    total electric field amplitude $E_0$ is fixed by the ionization
    yield $\rho^{\rm{max}}=2.7\times10^{18}$~cm$^{-3}$. White crosses
    in (a)-(c) indicate the values of the sawtooth
    waveform. (d) $N$-color
    sawtooth THz yield normalized to the 2-color pulse one
    when $\rho^{\rm max}$ is fixed. For comparison, the inset shows the
    THz yield for fixed pump energy flux $E_0^2\sum_k a_k^2$.}
\label{fig2}
\end{figure}

Figure~\ref{fig2} summarizes results from the LC model~\refeqs{eq:rho}{eq:tunnel} for Gaussian multi-color pulses with ionization yield $\rho^{\rm{max}}$ fixed to 10\% of $\rho_{0}$. For given $a_k$ and $\phi_k$ the field amplitude $E_0$ is chosen such that $\rho^{\rm{max}}=2.7\times10^{18}$~cm$^{-3}$. First, we fix relative amplitude and phase for the fundamental frequency to $a_1=1$, $\phi_1=-\pi/2$. This choice is arbitrary, because for multi-cycle pulses ($\tau=34$~fs) carrier-envelope phase effects are negligible. For $N=4$ colors, we are then left with six free parameters, $a_2,a_3,a_4$ and $\phi_2,\phi_3,\phi_4$. Because we cannot visualize the dependency of the THz yield on all six parameters in the same figure, we treat two-, three-,  and four-color cases separately and vary relative amplitude and phase of the highest harmonic only [see Figs.~\ref{fig2}(a)-(c)]. Clearly, we observe maximum THz yield for the sawtooth waveform in all cases. We checked that this behavior does not change when one selects the ionization yield to 5\% or 50\% of $\rho_0$.

An interesting issue is how the overall
 THz signal
depends on the number of harmonics that approximates the sawtooth
shape. We clarify this question in \reffig{fig2}(d). One can
see that the THz yield significantly increases till $N \sim 5$, while its quasi-linear growth
  saturates for larger $N$. This behavior is also
  supported analytically in the supplemental
  material. There, we also justify that  the pump wave shape which
  optimizes the THz yield approaches the sawtooth one at large $N$.  In the inset of \reffig{fig2}(d) we find it instructive to present the
  efficiency of the $N$-color sawtooth approximation in LC limit when
  the pump energy flux $E_0^2\sum_{k} a_k^2$ is fixed, instead of preserving the same ionization level. In this case, the sawtooth
  shape achieves a more impressive conversion up
  to a factor of $50$ because not only $v_f(t_n)$ but also $\rho_n$ grow
  considerably. Remarkably, somewhat
    similar wave shapes were found to increase the yield and electron
    recollision energy in high-order harmonic generation
    process by up to two orders of magnitude \cite{chipperfield09,haessler14}.

The advantage of the four-color approximation of a sawtooth
field is now studied using the unidirectional pulse propagation
equation that takes into account propagation effects in full space and time resolved
geometry. This 3D model was successfully tested against experimental data for THz generation from two-color pulses~\cite{Babushkin:prl:105:053903}. 
We use an adapted version of the unidirectional pulse propagation equation~\cite{Kolesik:pre:70:036604} 
for linearly polarized pulses
\begin{equation}
  \partial _{z}\widehat{E}=i\sqrt{k^2(\omega )-k_{x}^{2}-k_{y}^{2}}\widehat{E}+i\frac{\mu _{0}\omega ^{2}}{2k(\omega )} \widehat P_{\mathrm{NL}}.
\label{Eq.(1)}  
\end{equation}
Here, $\widehat{E}(k_{x},k_{y},z,\omega )$ is the pulse electric field expressed in Fourier domain with respect to transverse coordinates and time, $k=\omega n(\omega )/c$ is the wave number, $c$ is
the speed of light and $n(\omega)$ is the linear refractive index of
argon~\cite{Dalgarno:procrsoca:259:424}.  The nonlinear polarization
$\widehat{P}_{\mathrm{NL}}(\omega)=\widehat{P}_{\mathrm{Kerr}}(\omega)+i\widehat{J}(\omega)/\omega
+i\widehat{J}_{\mathrm{loss}}(\omega)/\omega $ accounts for third-order
nonlinear polarization $P_{\mathrm{Kerr}}(t)$, electron current $J(t)$
and a loss term $J_{\mathrm{loss}}(t)$ due to photon absorption during
ionization. The plasma current $J(t)$ is described by~\refeqs{eq:rho}{eq:tunnel}.
Since 3D propagation affects
relative phases, local intensities and pulse durations, we can
anticipate a reduced  THz conversion efficiency
 compared with the prediction of \reffig{fig2}(d). 

\begin{figure}
\includegraphics[width=\columnwidth]{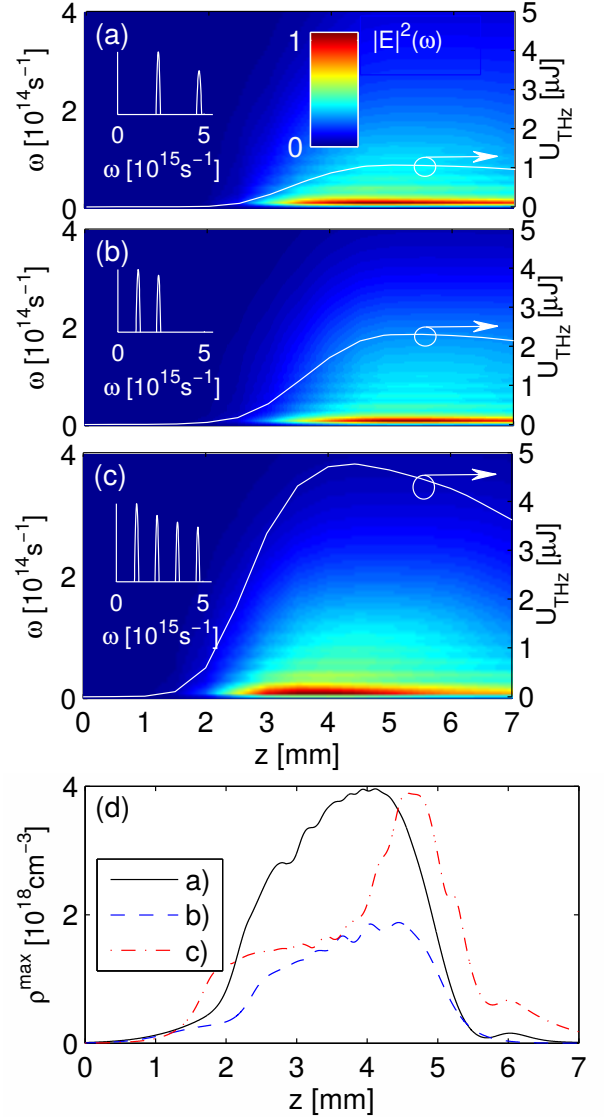} 
\caption{Low-frequency spectra (image plots) of 
(a) a two-color pulse for $\lambda_0=800$~nm and its second
harmonic in respective energy ratio of $\approx 0.06$, (b) a two-color pulse for $\lambda_0=1600$~nm and its second
harmonic in respective energy ratio of $\approx 0.4$, and (c) a four-color sawtooth pulse for $\lambda_0=1600$~nm 
(thus sum of harmonics energy in ratio $\approx
0.4$). The overall THz
energy ($\nu<100$~THz) in the numerical box, $U_{\rm{THz}}(z)$, versus
$z$ is shown by the right axes. 
The evolution of the peak free electron density for all three pulses
is shown in (d), revealing a comparable fraction ($\approx 10$\%) of the neutral atoms ionized.} 
\label{fig3}
\end{figure}

Figure~\ref{fig3}(a) displays the pulse spectrum (left axis) and the
THz yield below 100~THz (right axis) of a two-color 40-fs pulse with
overall energy of $300\,\mu$J. The fundamental wavelength is 800~nm
and 6\% of the pulse energy is in the second harmonic
(SH). The input beam width is $100\,\mu$m, and the pulse is
propagating along the longitudinal ($z$) axis focused over 5-mm focal
length in argon at 1 atm pressure. In this classical (non-optimized)
two-color setup the THz yield is about $1$~$\mu$J. Before passing over
to the four-color configuration in Fig.~\ref{fig3}(c), let us discuss
the second two-color configuration of Fig.~\ref{fig3}(b). Here we
shift the fundamental wavelength to 1600~nm and augment the SH ratio,
which leads to a roughly twofold increase of the THz yield. The SH
ratio is higher than that prescribed for the sawtooth waveform;
however, according to Fig.~\ref{fig2}(a) a too large coefficient $a_2$
has limited impact on the THz yield. It is worth noticing that, by
doubling the pump wavelength, only a factor
two is achieved in the enhancement of the THz yield. 
This departs from
the order of magnitude increase reported from setups  with different focusing conditions in
Refs. \cite{berge:prl:110:073901,Clerici:prl:110:253901}, but remains
consistent with the non-monotonic increase of the THz field strength
at high optical intensities ($>10^{14}$ W/cm$^2$) predicted in
\cite{Debayle:oe:22:13691}. 
In Fig.~\ref{fig3}(c) the results for the four-color configuration are
presented with pump pulses up to the fourth harmonic according to the
sawtooth waveform.  The THz yield is now $5$~$\mu$J [see
Fig.~\ref{fig3}(c)], i.e., 2.5 to 5 times larger than with the
two-color cases in Figs.~\ref{fig3}(a,b), which agrees with the LC
predictions presented in Fig.~\ref{fig2}. Finally, the electron
density evolution for the three pulses of Figs.~\ref{fig3}(a)-(c) is
shown in Fig.~\ref{fig3}(d).  Because pulse energy and focusing
conditions are the same, the plasma densities exhibit similar
dynamics.  In particular, a comparable fraction $\approx 10$\% of the
neutral atoms is ionized in all three cases, justifying the earlier
assumption of a constant ionization yield in the local current model.

In conclusion, THz energy depends not only on
  the number of electrons ionized, but also on the pump field
  waveform. We  have
identified the free electron velocity at the time instants of ionization
as being crucial for optimizing the THz yield. By exploiting this insight, we have shown that a sawtooth pump shape can boost THz conversion efficiency significantly compared to the standard two-color configuration.
Moreover, we provided arguments showing that in
  the regime of intensity clamping, which unavoidably appears when
  highest THz energy is targeted, the sawtooth shape is indeed
  optimal.  Comprehensive 3D simulations confirm this finding and
reveal 
an efficiency of 2\%,
which is unprecedented for THz generation in gases.

Numerical simulations were performed using high performance computing resources at Rechenzentrum Garching (RZG).
 AH acknowledges the support
of DFG (project HU
  1593/2-1). 
 IB is thankful to P.~Kinsler for useful
 discussions. 

\section{Supplemental Material}

\subsection{Expressions of $\rho(t)$, $\frac{d}{dt}J_B(t)$ and $\frac{d}{dt}J_A(t)$}

As preliminary computations we introduce the multi-color electric field Eq. (11) into Eq. (3) of the main article, assuming $|E(t)| \ll \beta$.
Ionization events occur at the instants $t_n$ of the electric field extrema. In the
neighborhood of $t_n$ the absolute value of the electric field is approximated by
\begin{equation}
\label{eq:Eparabolic}
|E(t)| \approx |E(t_n)| - \frac{1}{2}|\ddot{E}(t_n)|(t-t_n)^2,
\end{equation}
where $\ddot{E} \equiv \partial_t^2 E$. Following Ref. [17], the ionization rate can be expanded through Taylor series. Using \refeq{eq:Eparabolic}, it is easy to prove by recurrence that odd derivatives of $W$ are approximately zero
at $t=t_n$ and even derivatives of $W$ satisfy
\begin{widetext}
  \begin{equation}\label{eq:DmQST}
    \frac{d^{2m}}{dt^{2m}}{W}(|E(t_n)|) \approx (-1)^m (2m-1)!! W(|E(t_n)|) \left( \frac{\beta |\ddot{E}(t_n)|}{|E(t_n)|^2}\right)^m,\,\,\,m\geq 0,
  \end{equation}
\end{widetext}
where $p!!=p\cdot (p-2)\cdots3\cdot 1$ is the double factorial
of the odd integer number $p$ [$(-1)!!=1$].

By substituting \refeq{eq:DmQST} into the Taylor series of the
ionization rate, one finds 
\begin{widetext}
  \begin{equation*}
    \begin{split}
      \frac{W_n(t)}{W(|E(t_n)|)} & \approx
      \sum_{m=0}^{\infty}\left[(-1)^m (2m-1)!!  \left( \frac{\beta
            |\ddot{E}(t_n)|}{|E(t_n)|^2}\right)^m
        \frac{(t-t_n)^{2m}}{(2m)!} \right]\\
      & = \exp \left(- \frac{\beta |\ddot{E}(t_n)|}{2 |E(t_n)|^2}(t-t_n)^2\right),\\
    \end{split}
  \end{equation*}
\end{widetext}
since $(2m)!=(2m-1)!! m! 2^m$. The ionization rate
around the instant $t_n$ can thus be expressed as
\begin{equation}\label{eq:Wn}
W_n(t)\approx W(|E(t_n)|)\mathrm{e}^{-\left( \frac{t-t_n}{\tau_n}\right)^2},
\end{equation}
where $\tau_n$ is interpreted as a characteristic time-length of the ionization event defined by
\begin{equation}\label{eq:taun}
\frac{1}{\tau_n^2}=\frac{\beta |\ddot{E}(t_n)|}{2 |E(t_n)|^2}=
\frac{1}{3}E_a(U_i/U_H)^{3/2}\frac{|\ddot{E}(t_n)|}{|E(t_n)|^2}.
\end{equation}

Assuming that the superposition principle holds, we obtain the evaluation of the ionization rate
\begin{equation}\label{eq:W}
W(t)\approx \sum_n W(|E(t_n)|)\mathrm{e}^{-\left( \frac{t-t_n}{\tau_n}\right)^2}.
\end{equation}

\subsection{Free electron density}

For hydrogen-like atoms, the free electron density is given by Eq.  (1) of
the paper, yielding
\begin{equation}\label{eq:rho_s}
\rho(t)=\rho_0 \left[ 1 - \exp\left(-\int_{-\infty}^t W(\tau) \,d\tau \right)  \right],
\end{equation}
whenever $\rho(t\rightarrow-\infty)=0$. Employing \refeq{eq:W} leads to
\begin{equation}\label{eq:iW}
\begin{split}
\int_{-\infty}^t W(\tau) \,d\tau & \approx  \sum_n \left[ W(|E(t_n)|)  \int_{-\infty}^t \mathrm{e}^{-\left( \frac{t-t_n}{\tau_n}\right)^2} \,d\tau \right] \\
&  = \sum_n \left[ \sqrt{\pi} \tau_n W(|E(t_n)|)  H_n(t-t_n) \right], \\
\end{split}
\end{equation}
where $H_n(t)=\frac{1}{2}(1+\erf(t/\tau_n))$ tends to the standard Heaviside step function $\Theta(t)$ when $\tau_n\rightarrow 0$, i.e.,
\begin{equation}\label{eq:limW}
\lim_{\tau_n\rightarrow 0} H_n(t) = \Theta(t) = \left\lbrace \begin{array}{ll}
1, & \text{if }t\geq 0, \\
0, & \text{if }t< 0. \\
\end{array} \right.
\end{equation}

To evaluate $\rho(t)$ [\refeq{eq:rho_s}] we exploit the limit given by \refeq{eq:limW}.
First, let us consider a single ionization event.
We substitute \refeq{eq:iW} for $n=1$ into \refeq{eq:rho_s} using $H_1(t-t_1) \simeq \Theta(t-t_1)$:
\begin{equation}\label{eq:rho1}
\begin{split}
\rho(t)&=\rho_0 \left( 1 - \mathrm{e}^{- \sqrt{\pi} \tau_1 W(|E(t_1)|)  H_1(t-t_1)}  \right) \\
&\approx  \rho_0 \left( 1 - \mathrm{e}^{- \sqrt{\pi} \tau_1 W(|E(t_1)|)  \Theta(t-t_1)}  \right)\\
&= \rho_0 \left( 1 - \mathrm{e}^{- \sqrt{\pi} \tau_1 W(|E(t_1)|)}  \right) \Theta(t-t_1) \approx  \delta\rho_1 H_1(t-t_1), \\
\end{split}
\end{equation}
where $\delta\rho_1$ denotes the density jump of the first ionization event. Second, for two ionization events a similar reasoning yields
\begin{widetext}
  \begin{equation}\label{eq:rho2}
    \begin{split}
      \rho(t)&= \rho_0 \left( 1 - \mathrm{e}^{- \sqrt{\pi} \tau_1 W(|E(t_1)|)  H_1(t-t_1)-\sqrt{\pi} \tau_2 W(|E(t_2)|)  H_2(t-t_2)}  \right) \\
      & \approx \rho_0 \left( 1 - \mathrm{e}^{- \sqrt{\pi} \tau_1 W(|E(t_1)|)  \Theta(t-t_1)- \sqrt{\pi} \tau_2 W(|E(t_2)|)  \Theta(t-t_2)}  \right)\\
      & = \delta\rho_1\Theta(t-t_1)+ \rho_0 \left( 1 - \mathrm{e}^{- \sqrt{\pi} \tau_1 W(|E(t_1)|)- \sqrt{\pi} \tau_2 W(|E(t_2)|)}  - \frac{\delta\rho_1}{\rho_0}\right) \Theta(t-t_2) \\
      & \approx \delta\rho_1 H_1(t-t_1)+\delta\rho_2 H_2(t-t_1), \\
    \end{split}
  \end{equation}
\end{widetext}
where $\delta\rho_2 / \rho_0=\mathrm{e}^{- \sqrt{\pi} \tau_1 W(|E(t_1)|)} - 
\mathrm{e}^{- \sqrt{\pi} \tau_1 W(|E(t_1)|)- \sqrt{\pi} \tau_2 W(|E(t_2)|)}$ is
the density jump of second ionization event. By a recursive reasoning, we find the following step-wise model for the
density
\begin{equation}\label{eq:rhon_s}
\rho(t)= \sum_n \delta\rho_n H_n(t-t_n),
\end{equation}
where the density jumps for $n> 1$ are defined by
\begin{equation}\label{eq:deltarhon}
 \delta\rho_n = \rho_0 \mathrm{e}^{- \sum_{j=1}^{n-1} \sqrt{\pi} \tau_j W(|E(t_j)|)}\left(1- \mathrm{e}^{- \sqrt{\pi} \tau_n W(|E(t_n)|)} \right).
\end{equation}

The free electron density results from the
superposition of all step-like ionization events. Equation (\ref{eq:deltarhon}) guarantees
that $\rho(t)$ will saturate for high intensities and/or for long laser pulse durations.

\subsection{THz contribution from ${\bf \frac{d}{dt}J_B(t)}$}

Following Ref. [17] of the main article, $J(t)$ expresses as
\begin{equation}
\label{eq:Jn}
J(t)= J_A(t) + J_B(t),
\end{equation}
\begin{equation}\label{eq:JnA}
J_A(t)= q\sum_n \left[ \delta\rho_n  v_f(t)  H_n(t-t_n) \right],
\end{equation}
\begin{equation}\label{eq:JnB}
J_B(t)= - q\sum_n \left[ \delta\rho_n v_f(t_n) \mathrm{e}^{-\frac{t-t_n}{\tau_c}}   
H_n(t-t_n) \right].
\end{equation}

$J_A(t)$ is expected to dominate in the high-frequency part of
the spectrum, although for some pulse configurations it may also affect the THz band.
In contrast, the component
$J_B(t)$ contributes mostly to low frequencies. 
According to the LC photo-current model, the field radiated
by accelerated electrons is proportional to the derivative of the current. We thus take the time-derivative of \refeq{eq:JnB} and then apply a Fourier transform, yielding
\begin{widetext}
  \begin{equation}\label{eq:FdJnB}
    \mathcal{F}\left(\partial_t{J}_B(t)\right)= - q\sum_n\left[   \frac{\delta\rho_n v_f(t_n)}{\sqrt{\pi}\tau_n}
      \mathcal{F}\left( \mathrm{e}^{-\frac{t-t_n}{\tau_c}-\left(\frac{t-t_n}{\tau_n}\right)^2}\right) - \frac{\delta\rho_n v_f(t_n)}{\tau_c}\mathcal{F}\left( \mathrm{e}^{-\frac{t-t_n}{\tau_c}}H_n(t-t_n)\right) \right],
  \end{equation}
\end{widetext}
where, by definition, $\mathcal{F}\left(f(t)\right)=\frac{1}{\sqrt{2\pi}}\int_{-\infty}^{+\infty} f(t) \mathrm{e}^{i\omega t} \,dt$.

The first Fourier transform in the right-hand side (RHS) of \refeq{eq:FdJnB} is
\begin{widetext}
  \begin{equation}
    \label{fourier}
    \mathcal{F}\left( \mathrm{e}^{-\frac{t-t_n}{\tau_c}-\left(\frac{t-t_n}{\tau_n}\right)^2}\right) = \frac{\tau_n}{\sqrt{2}} \mathrm{e}^{i t_n \omega - \frac{1}{4}(\tau_n^2\omega^2 - \frac{\tau_n^2}{\tau_c^2}+
      \frac{2 i \tau_n^2}{\tau_c}\omega)} \approx \frac{\tau_n}{\sqrt{2}} \mathrm{e}^{i t_n \omega},
  \end{equation}
\end{widetext}
which we approximate in the THz-band ($\omega \rightarrow 0$) using $\omega^2 \ll \omega$ and
$\tau_n \ll \tau_c$. THz generation due to this component proceeds from constructive interference of contributions in $\mathrm{e}^{i t_n \omega}$. 

We filter this component in the interval $[0, \omega_{\rm co}]$, where $\omega_{\rm co}$ is the cut-off frequency in the THz band, by using the rectangular filter $\Pi(\omega)=\Theta(\omega+\omega_{\rm co})\left(1- \Theta(\omega-\omega_{\rm co})\right)$, such that
\begin{equation}\label{eq:sincdJnB}
\mathcal{F}^{-1}\left(  \frac{\tau_n}{\sqrt{2}} \mathrm{e}^{i t_n \omega} \Pi(\omega) \right)=
\frac{\tau_n \omega_{\rm co}}{\sqrt{\pi}}\sinc\left(\omega_{\rm co}(t-t_n)  \right).
\end{equation}

The second Fourier transform in the RHS of \refeq{eq:FdJnB} expresses as
\begin{widetext}
  \begin{equation}
    \mathcal{F}\left( \mathrm{e}^{-\frac{t-t_n}{\tau_c}}H_n(t-t_n)\right)=
    \frac{i \tau_c  \mathrm{e}^{i\omega t_n -\frac{\tau_n^2(i+\tau_c
          \omega)^2}{4\tau_c^2} }  }{\sqrt{2\pi}\left(i+\tau_c \omega
      \right)} \approx \frac{i\tau_ce^{i\omega
        t_n}}{\sqrt{2\pi} (i+\tau_c\omega)},
    \label{eq:jbsec}
  \end{equation}
\end{widetext}
which is a peak function lying, in the case of $\omega_{\rm co}\gg1/\tau_c$, already in the low frequency domain and will thus not be filtered.
\refeq{eq:jbsec}, together with \refeq{fourier} gives Eq.~(10a) of the main paper.

The THz component of $\partial_t{J}_B(t)$ reads as
\begin{widetext}
  \begin{equation}\label{eq:THzdJnB}
    \partial_t{J}^{\rm THz}_B(t)= -q \sum_n  \left[\delta\rho_n v_f(t_n) \left( \frac{\omega_{\rm co} \sinc\left(\omega_{\rm co}(t-t_n)  \right)}{\pi}-  \frac{\mathrm{e}^{-\frac{t-t_n}{\tau_c}}H_n(t-t_n)}{\tau_c} \right) \right].
  \end{equation}
\end{widetext}

\subsection{THz contribution from ${\bf \frac{d}{dt}J_A(t)}$}

Due to the complexity of its expression, we propose an approximate model of $\partial_t{J}^{\rm THz}_A(t)$ based on Taylor
series and the following approximations:
\begin{enumerate}
\item For technical convenience, the slowly-varying envelopes $\mathcal{E}_k(t)$
in Eq. (11) of the main article are square cosinus with compact support:
\begin{equation}\label{eq:envapprox}
\mathcal{E}_k(t) \approx E_0 \cos^2\left(\frac{\pi t}{2 \tau_k}  \right)
\Theta(t+\tau_k)\left(1- \Theta(t-\tau_k)\right),
\end{equation}
where $\tau_k$ denotes the duration of the $k$th color.
\item The pulse has a large number of cycles:
\begin{equation}\label{eq:slowvar}
\frac{1}{\omega_k}\ll \tau_k.
\end{equation}
\item We simplify more the free electron density [\refeq{eq:rhon_s}] and consider standard Heaviside step functions:
\begin{equation}\label{eq:rhonapprox}
\rho(t)= \sum_n \delta\rho_n \Theta(t-t_n).
\end{equation}
\end{enumerate}

By integrating by parts Eq. (8) of the paper and neglecting time-derivatives of the envelope, we get
\begin{equation}\label{eq:vfapprox}
v_f(t) \approx \frac{q}{m} \sum_{k=1}^N a_k
\mathcal{E}_k(t)
\frac{\sin(k \omega_0 t + (-1)^k \phi_k)}{k \omega_0}.
\end{equation}

We take the Fourier transform of \refeq{eq:JnA} in which we apply Eqs. (\ref{eq:rhonapprox}) and (\ref{eq:vfapprox}) with the
envelope \refeq{eq:envapprox} to obtain
\begin{widetext}
  \begin{equation*}
    \begin{split}
      \mathcal{F}\left(J_A(t)\right)=& q\sum_n \left[ \frac{\delta\rho_n}{\sqrt{2\pi}} \int_{-\infty}^{+\infty}  v_f(t)  H_n(t-t_n)  \mathrm{e}^{i\omega t} \,dt  \right]  \\
      & \approx q\sum_n \left[ \frac{\delta\rho_n}{\sqrt{2\pi}} \int_{t_n}^{+\infty}  v_f(t) \mathrm{e}^{i\omega t} \,dt  \right] \\
      & \approx \frac{q^2E_0}{m}\sum_n\sum_k \left[\frac{\delta\rho_na_k}{\sqrt{2\pi}k \omega_0} \times \right. \\
      &  \int_{t_n}^{+\infty} \cos^2\left(\frac{\pi t}{2\tau_k}  \right) \Theta(t+\tau_k)\left(1- \Theta(t-\tau_k)\right) \times \\
      & \left. \sin(k \omega_0 t + (-1)^k \phi_k) \mathrm{e}^{i\omega t} \,dt  \right] = \frac{q^2E_0}{m} \sum_n\sum_k \left[ \frac{\delta\rho_na_k}{\sqrt{2\pi}k \omega_0} \times \right. \\
      & \left. \int_{t_n}^{\tau_k} \cos^2\left(\frac{\pi t}{2\tau_k}  \right) \sin(k \omega_0 t + (-1)^k \phi_k) \mathrm{e}^{i\omega t} \,dt  \right]. \\
    \end{split}
  \end{equation*}
\end{widetext}
For practical use, we only consider
the first two terms
in the Taylor expansion of the previous expression around $\omega=0$. After cumbersome calculations (not detailed),
we get
\begin{equation}\label{eq:wJnA}
\mathcal{F}\left(\partial_t{J}_A(t)\right)=-i\omega \mathcal{F}\left(J_A(t)\right)\approx -i\omega \tilde{C}_0+\omega^2 \tilde{C}_1,
\end{equation}
\begin{equation}
\tilde{C}_0=\frac{q^2}{\sqrt{2\pi}m}\sum_n\sum_k \left[ \delta\rho_n \frac{E_k(t_n)}{k^2 \omega_0^2} \right],
\end{equation}
\begin{equation}
\tilde{C}_1=\frac{q^2}{\sqrt{2\pi}m}\sum_n\sum_k \left[ \delta\rho_n t_n \frac{E_k(t_n)}{k^2 \omega_0^2} \right].
\end{equation}
This approximation is applicable if the low-frequency range we
  consider is small compared to $\omega_0$, that is,
  $\omega_{\rm co}\ll \omega_0$.
Finally, we filter out \refeq{eq:wJnA} using the cut-off frequency $\omega_{\rm co}$ to obtain
\begin{equation}\label{eq:THzA}
\partial_t{J}^{\rm THz}_A(t)\approx C_0\omega_{\rm co}^2 \dsinc(\omega_{\rm co} t) + C_1\omega_{\rm co}^3 \ddsinc(\omega_{\rm co} t),
\end{equation}
\begin{equation}\label{eq:C0}
C_0=\frac{q^2}{\pi m}\sum_n\sum_k \left[ \delta\rho_n \frac{E_k(t_n)}{k^2 \omega_0^2} \right],
\end{equation}
\begin{equation}
C_1=-\frac{q^2}{\pi m}\sum_n\sum_k \left[ \delta\rho_n t_n \frac{E_k(t_n)}{k^2 \omega_0^2} \right],
\end{equation}
where $\dsinc(x)$ and $\ddsinc(x)$ denote the first two successive
$x$-derivatives of the function $\sinc(x)$. Equation (10b) of the main
paper follows from \refeq{eq:wJnA} and the  inequality $|C_1\omega_{\rm co}^3| \ll |C_0\omega_{\rm co}^2|$.

\subsection{Growth of the THz energy with $N$ for fixed ionization degree}

The $N$-color approximation of the sawtooth waveform without envelope is given by
\begin{equation}\label{eq:EN}
E(t) =a_N \sum_{k=1}^N \frac{(-1)^{k+1}}{k}\sin (k\omega_0 t),
\end{equation}
where $a_N$ is adjusted to the desired ionization degree. The absolute extrema of Eq. (\ref{eq:EN}) are attained at the instants
\begin{equation}
\omega_0 t_n = \pm \frac{N}{N+1}\pi + 2\pi n.
\end{equation}

After easy manipulations assuming $\tau_c \rightarrow \infty$ for simplicity, the free electron velocity at $t=t_n$ reads as
\begin{equation}
\label{eq:vfNn}
v_f^N(t_n) \propto a_N \chi(N) \equiv a_N \sum_{k=1}^N \frac{(-1)^k}{k^2} \cos \left(\frac{N}{N+1}k\pi\right).
\end{equation}
Similarly, the value of $E(t_n)$ only depends on the number of color
$N$ and so do the maximum ionization rate $W(t_n)$ and the density
jumps $\delta \rho_n$. 
Besides, from 
  Eqs. (\ref{eq:FdJnB}) and (\ref{eq:wJnA}), one has in Fourier
  domain
  \begin{widetext}
    \begin{equation}
      \label{eq:8}
      \mathcal{F}[\partial_t{J}](\omega) \approx
      \frac{-q}{\sqrt{2\pi}}\sum_n\delta\rho_n
      v_f(t_n)e^{it_n\omega}\frac{\omega}{i/\tau_c+\omega}
      -i\tilde C_0 \omega +  \tilde C_1 \omega^2. 
    \end{equation}
  \end{widetext}
For $\omega \gtrapprox 1/\tau_c$, every of the first $N$
  terms has approximately constant amplitude, whereas the last two
  terms are $\sim O(\omega)$ and thus typically negligible in THz
  range. Thus, to some approximation we can write

\begin{equation}
\label{eq:16}
\mathcal{F}[\partial_t{J}_{\rm THz}](\omega) \propto \sum_n \delta \rho_n v_f(t_n) e^{i\omega t_n} \equiv B(\omega).
\end{equation}
Furthermore, $\int_0^{\omega_{co}} BB^*d\omega$
should give the emitted THz energy flux $U_{\rm THz}$ up to a
constant factor. For short enough pulses, we
assume that the inverse of time intervals $1/(t_n-t_m)$ lies outside
the THz range whatever $n,m$ may be, and thus
\begin{equation}
\label{eq:13}
U_{\rm THz} \propto \left(\sum_n \delta \rho_n v_f(t_n) \right)^2.
\end{equation}

\begin{figure}
\begin{centering}
\includegraphics[width=\columnwidth]{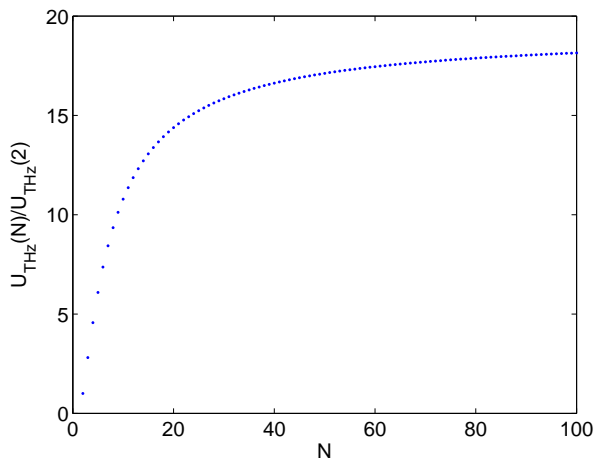}
\caption{Growth of the THz energy with the number of harmonics.} 
  \label{fig:UNU2}
\end{centering}
\end{figure}

Accounting for the pulse envelope, $U_{\rm THz}$ is mostly yielded by the contribution, $U_{\rm THz}(N)$, when \refeq{eq:EN} is maximum close to the instants $t = \pm N\pi/[(N+1)\omega_0]$ over the sawtooth period. By taking into account Eq. (\ref{eq:vfNn}) and the hypothesis of constant ionization level, $\sum_n \delta\rho_n = \text{constant}$, we obtain $U_{\rm THz}(N) \propto v_f^N(t_n)^2$. To preserve constant ionization, the amplitude $a_N$ must smoothly decrease with the number of colors $N$. Therefore, using $\chi^2(2)=9/64$, we can bound from above the growth of the THz energy as 
\begin{widetext}
  \begin{equation}
    \label{eq:UNU2}
    \frac{U_{\rm THz}(N)}{U_{\rm THz}(2)} \leq  \frac{64}{9}\chi^2(N) = \frac{64}{9} \left(\sum_{k=1}^N \frac{(-1)^k}{k^2} \cos \left(\frac{N}{N+1}k\pi\right) \right)^2.
  \end{equation}
\end{widetext}
Equation (\ref{eq:UNU2}) is plotted in Fig. \ref{fig:UNU2}. Due to the assumptions done, the values are larger than those of Fig. 2(d) in the main paper. Equation (\ref{eq:UNU2}), however, justifies the saturation in the THz yields reported by the latter figure.

\subsection{Optimality of sawtooth-like waveforms}

In this section we consider another
approach to justify the optimality of the sawtooth shape.
We consider a
  single cycle of the pump field oscillation and assume a single
  ionization event there. We optimize
  $v_f(t_n)$ by varying the wave shape. For technical convenience, we keep both the whole pump intensity and $\delta \rho_n$ constant. We also apply the limit of an infinite decay time $\tau_c \rightarrow +\infty$.
The ``input'' for our optimization procedure
is the pump waveform
\begin{equation}
  \label{eq:1}
  E(t) = E_0 \sum_{k=1}^{N} a_k \cos(k \omega_0 t + \phi_k),
\end{equation}
where $\omega_k \equiv k\omega_0$ is the $k$th harmonic of the
fundamental frequency $\omega_0$. \refeq{eq:1}
  is obtained from Eq.~(11) of the main article assuming constant envelope amplitudes  $\mathcal{E}_k(t)
\equiv E_0$.
Importantly, we consider waveforms having the same average intensity
over the pump period:
\begin{equation}
  \label{eq:3}
I=  \int_{-T_0/2}^{T_0/2} E^2(\tau) d\tau \propto \sum_k a_k^2 = 1.
\end{equation}
The exact value of the constant on the right hand side is not
important, since we optimize the wave shape and not its intensity value.
Without loss of generality we assume an extremum of
$E(t)$ at $t=0$ such that
\begin{equation}
  \label{eq:6}
  \partial_t E(0) = - \sum_k  k a_k \omega_0 \sin{\phi_k} = 0.
\end{equation}
The quantity to be maximized is then
\begin{equation}
  \label{eq:5}
  v_f(0) = \frac{q}{m} \sum_k \frac{a_k}{k\omega_0}\sin{\phi_k},
\end{equation}
after choosing the initial time $t_0$ such that $\sum_k \frac{a_k}{k\omega_0}\sin{(k\omega_0 t_0+\phi_k)} = 0$.

It is then evident that, since $v_f(0)$ is linear in $\sin{\phi_k}$ and assuming all $a_k$'s having the same sign, the extrema of \refeq{eq:5} are achieved at the phase values $\phi_k=\pm \pi/2$, as expected for the sawtooth phases.

Furthermore, we can rewrite the optimization problem \refeq{eq:5} with
  constraints \refeqs{eq:3}{eq:6}  in a vectorial form. We introduce the vectors $\vect
  a=\{a_1,a_2,\ldots,a_N\}$, $\vect b= \omega_0 \{1,2,\ldots,N\}$ and $\vect c=(1/\omega_0) \{1,1/2,\ldots,1/N\}$.
In the limit of large $N$, we determine the extremum of $(\vect c,\vect a)$, where $(.,.)$ mean
scalar multiplication of two vectors, assuming $(\vect b,\vect a)=0$
and $|\vect a|=1$.

The optimum of such a problem is achieved on the vector $\vect a$ defined by
\begin{equation}
  \label{eq:10}
  \vect a = C\left(\vect c -\vect b \frac{(\vect b,\vect c)}{(\vect b,
    \vect b)}\right),
\end{equation}
where the constant $C$ is selected to assure $|\vect a|=1$, while $(\vect b, \vect c) = N$ and $|\vect b|^2 \equiv \omega_0^2 S$ with
\begin{equation}
\label{b2}
S =\sum_{k=1}^N k^2 = N(N+1)(2N+1)/6.
\end{equation}
With the previous relations we get 
\begin{equation}
  \label{eq:11}
  a_k = \frac{C}{\omega_0} \left(\frac1k -\frac{k N}{S}\right).
\end{equation}
Moreover, one has $|\vect a|^2 = 1 = C (\vect c, \vect a)$ and in the limit of large $N$ it is straightforward to obtain 
\begin{equation}
  \label{eq:15}
C \simeq \frac{(\vect c,\vect a) \omega_0^2}{\sum_k k^{-2}} \equiv
\frac{\omega_0^2}{C \sum_k k^{-2}},
\end{equation}
yielding at leading order
\begin{equation}
\label{ak}
a_k \simeq \frac{1}{k \sqrt{\sum_{j=1}^N j^{-2}}},
\end{equation} 
which is the right normalization of the sawtooth amplitude.

For finite $N$ it is easy to show that \refeq{eq:11} approaches the
 $N$-truncation of the ideal sawtooth rapidly as $N$ increases.
Indeed, from Eq. (\ref{eq:11}) the deviation in amplitude from the ideal sawtooth profile is yielded by $\delta a_k =-kN/S$, i.e.,
\begin{equation}
  \label{eq:12}
|\vect \delta \vect a|^2 = \frac{N^2}{S^2}\sum_{k=1}^N k^2=\frac{N^2}{S}.
\end{equation}
With $S\sim N^3$, $|\vect \delta \vect a|^2\sim 1/N$ vanishes as $N$
increases. That is, in this limit,
$a_k\to \mbox{const}/k$, which means that
for large enough number of harmonics the optimal wave shape
approaches the sawtooth one.


%

\end{document}